\documentclass{desyproc}
\usepackage{upgreek}

\begin{document}

\title{Search for hidden-photon Dark Matter with FUNK}

\author{
{\slshape
R.~Engel$^1$\footnote{\texttt{ralph.engel@kit.edu}},
D.~Veberi\v{c}$^1$,
C.~Sch\"afer$^1$,
A.~Andrianavalomahefa$^1$,
K.~Daumiller$^1$,
B.~D\"obrich$^2$,
J.~Jaeckel$^3$,
M.~Kowalski$^{4,5}$,
A.~Lindner$^4$,
H.-J.~Mathes$^1$,
J.~Redondo$^6$,
M.~Roth$^1$,
T.~Schwetz-Mangold$^1$,
and
R.~Ulrich$^1$
}
\\[1ex]
$^1$Institute for Nuclear Physics, Karlsruhe Institute of Technology (KIT), Germany\\
$^2$Physics Department, CERN, Geneva, Switzerland\\
$^3$Institute for Theoretical Physics, Heidelberg University, Germany\\
$^4$Deutsches Elektronen Synchrotron DESY, Hamburg, Germany\\
$^5$Department of Physics, Humboldt University, Berlin, Germany\\
$^6$Department of Theoretical Physics, University of Zaragoza, Spain
}

\contribID{Engel\_Ralph}

\confID{16884}
\desyproc{DESY-PROC-2017-XX}
\acronym{Patras 2017}
\doi

\maketitle

\begin{abstract}
It has been proposed that an additional U(1) sector of hidden photons could
account for the Dark Matter observed in the Universe. When passing through an
interface of materials with different dielectric properties, hidden photons can
give rise to photons whose wavelengths are related to the mass of the hidden
photons. In this contribution we report on measurements covering the visible
and near-UV spectrum that were done with a large, 14\,m$^2$ spherical metallic
mirror and discuss future dark-matter searches in the eV and sub-eV range by
application of different electromagnetic radiation detectors.
\end{abstract}

For the introduction to the hidden-photon physics and related extension of the
Standard Model see~\cite{Dobrich:2014kda,Dobrich:2015tpa}. For results of a
similar experiment, although with a smaller mirror
see~\cite{Suzuki:2015sza,Suzuki:2015vka}.

\section{Experimental setup}

\begin{figure}[t]
\def\figh{0.35}
\centering
\includegraphics[height=\figh\textwidth]{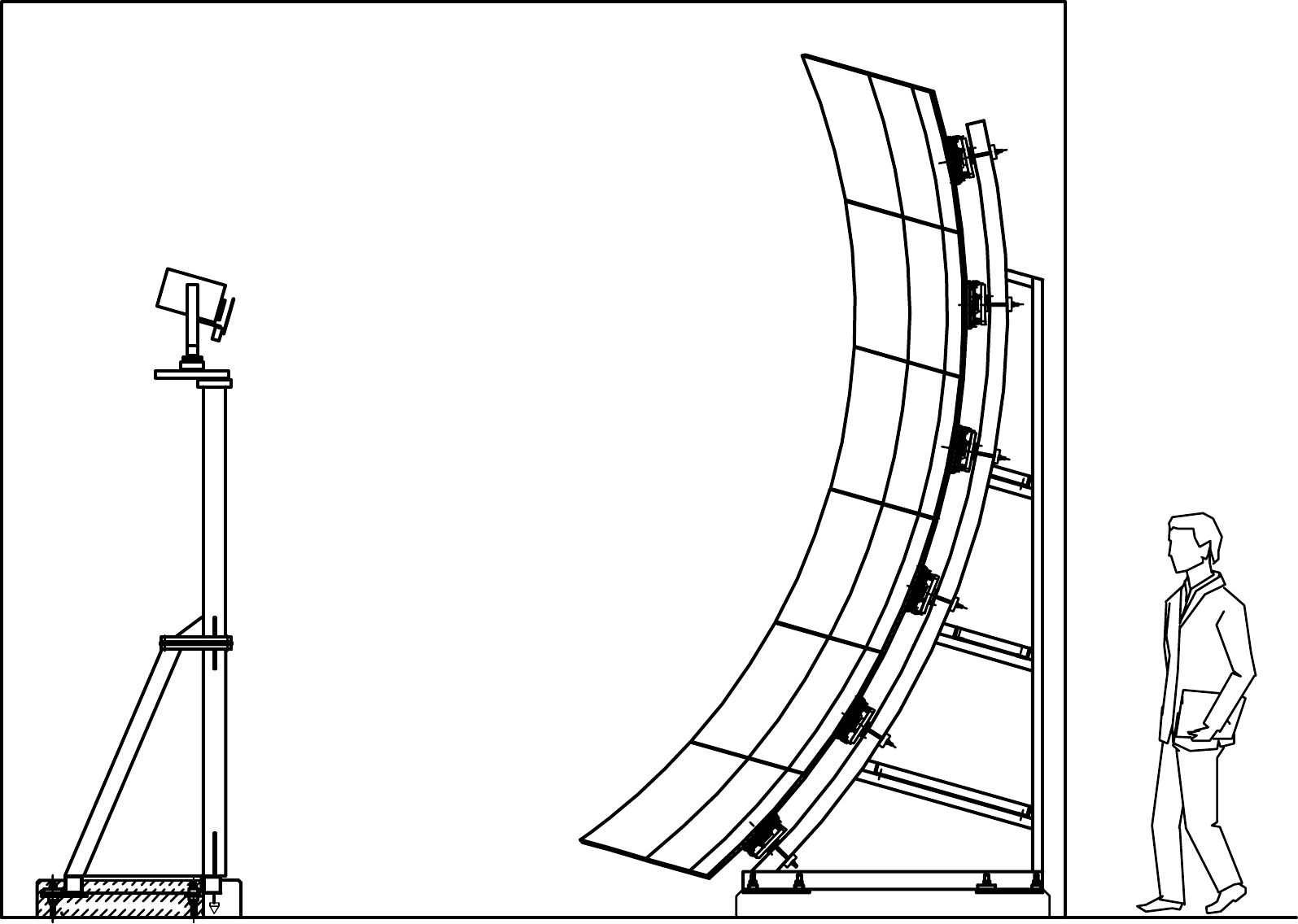}\hfill
\includegraphics[height=\figh\textwidth]{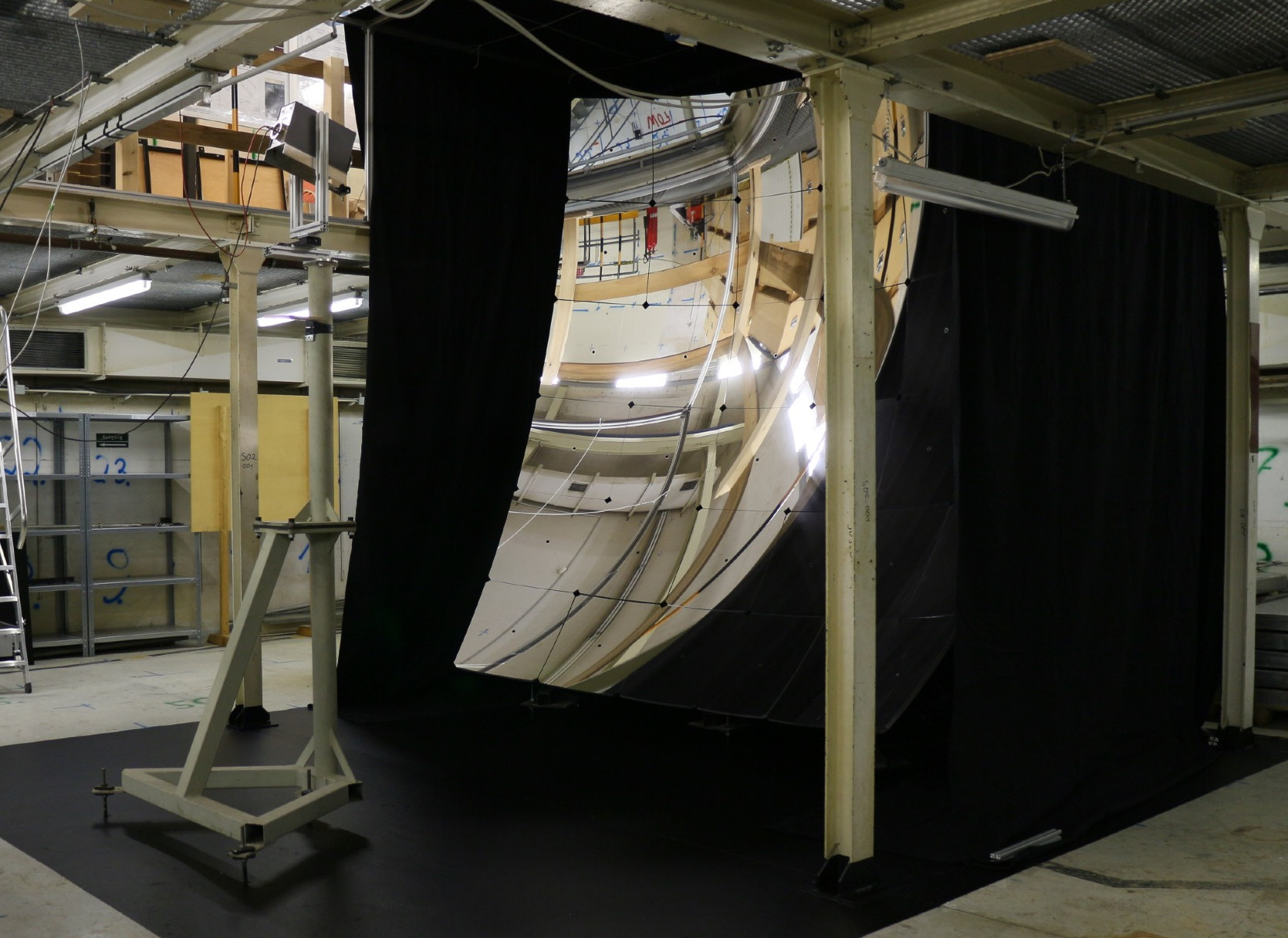}
\caption{Schematic (left) and photo (right) of the experimental setup.}
\label{f:setup}
\end{figure}

For this experiment a mirror composed of 36 segments is used. For more details
on the setup see~\cite{Veberic:2015yua,FunkICRC2017}. The experiment is set-up
in a light-tight window-less room with concrete walls of at least 2\,m
thickness. The inner area (see Fig.~\ref{f:setup}), encompassing the camera and
the mirror, is additionally light insulated with a thick black curtain and a
120\,$\upmu$m layer of opaque polyethylene foil.

As the light detector a 29\,mm diameter photomultiplier (PMT) ET 9107BQ with
very low dark-current properties was chosen.  The PMT has a blue-green
sensitive bialkali photocathode with the quantum efficiency extended into the
ultra-violet range with the peak quantum efficiency of around 28\% and
excellent single electron and pulse-height resolution, suitable for the photon
counting. The PMT camera is placed on a motorized linear-stage that can drive
it (perpendicularly to the mirror axis) in and out of the center of the
spherical mirror. The PMT front is additionally equipped with a motorized
shutter that can obscure the entrance of photons.

\begin{figure}[t]
\def\figh{0.372}
\centering
\includegraphics[height=\figh\textwidth]{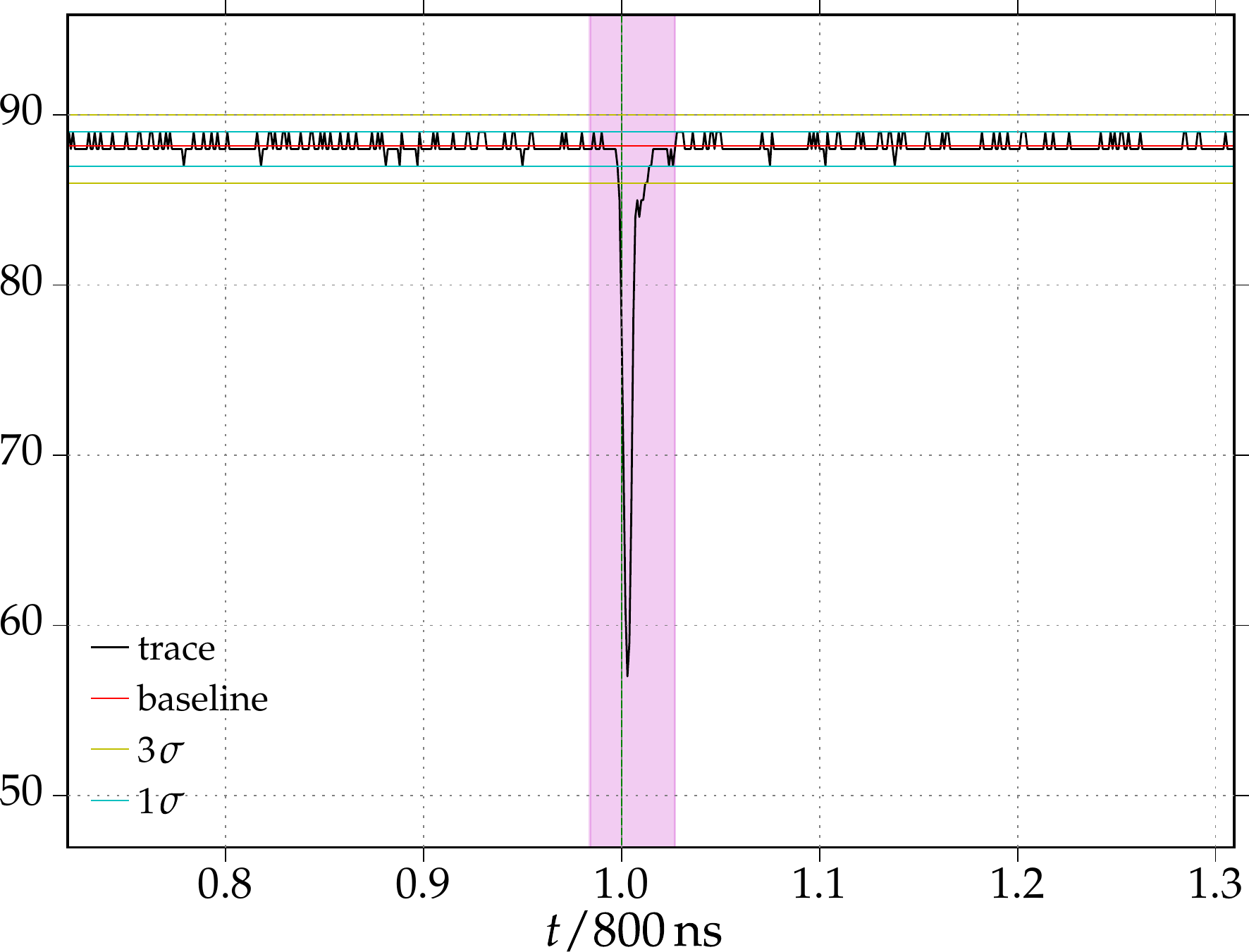}\hfill
\includegraphics[height=\figh\textwidth]{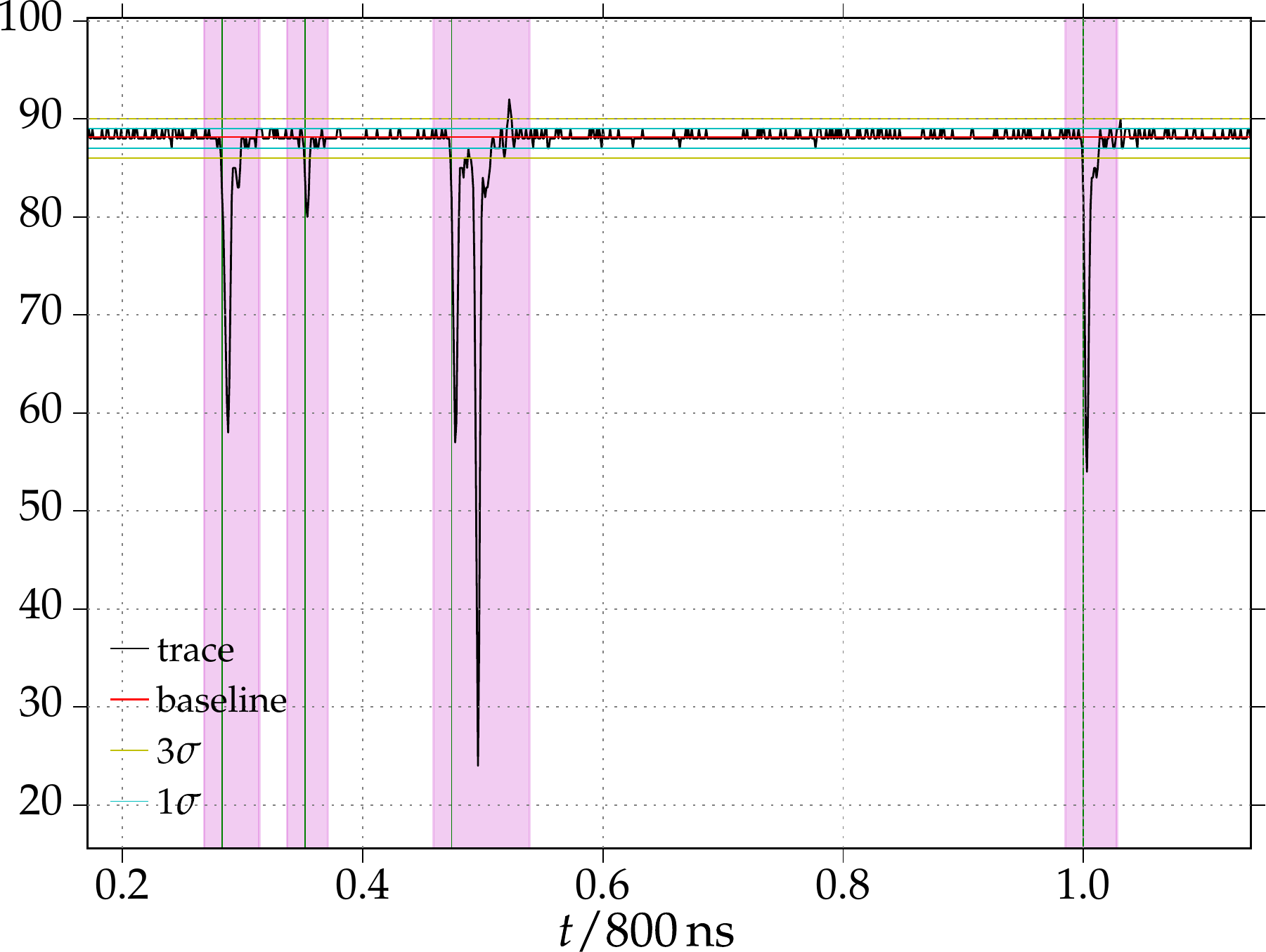}
\caption{Typical examples of captured traces with single pulse (left) and many
pulses within a short time span (right).}
\label{f:traces}
\end{figure}

\begin{figure}[t]
\def\figh{0.39}
\centering
\includegraphics[height=\figh\textwidth]{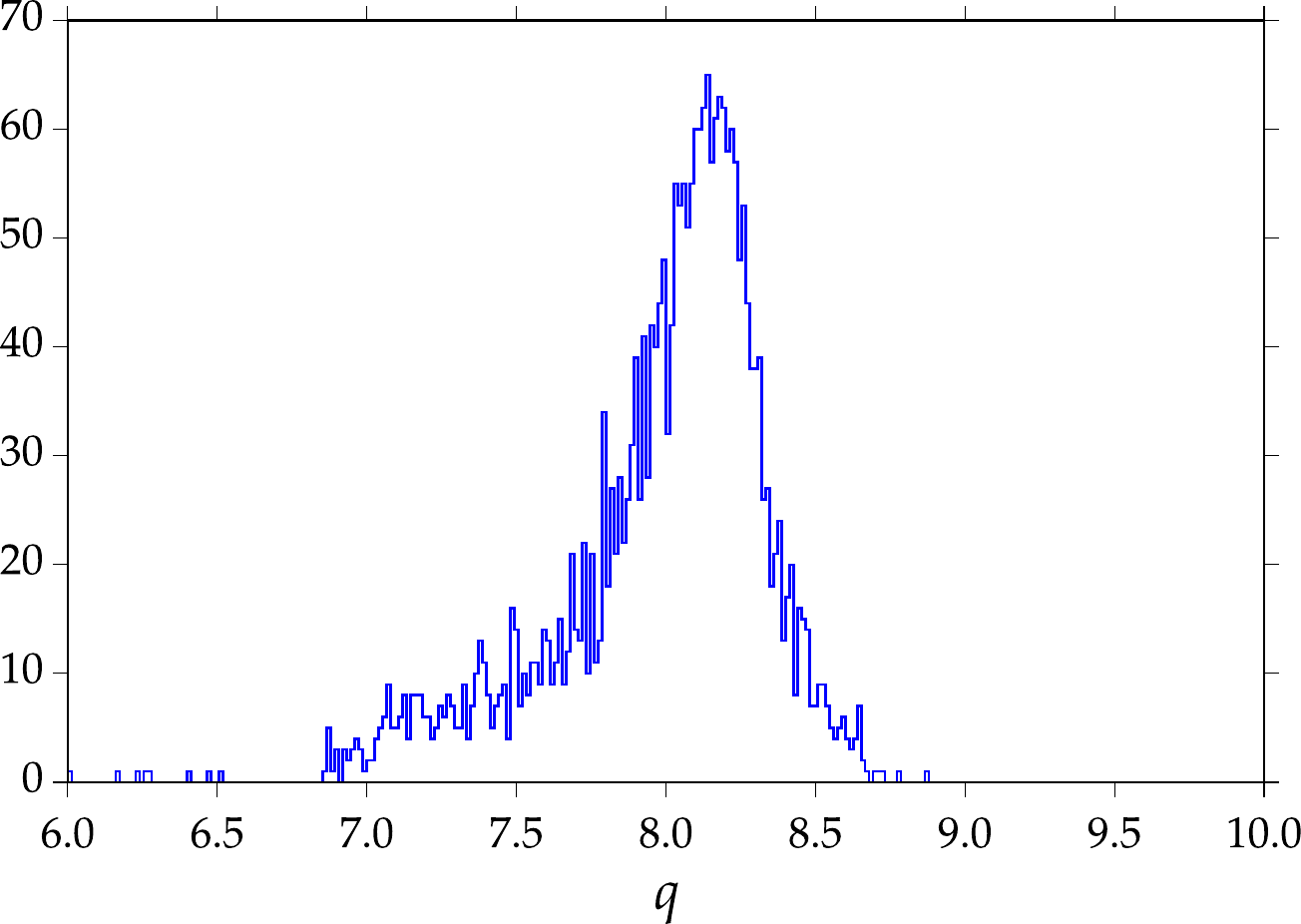}\hfill
\includegraphics[height=\figh\textwidth]{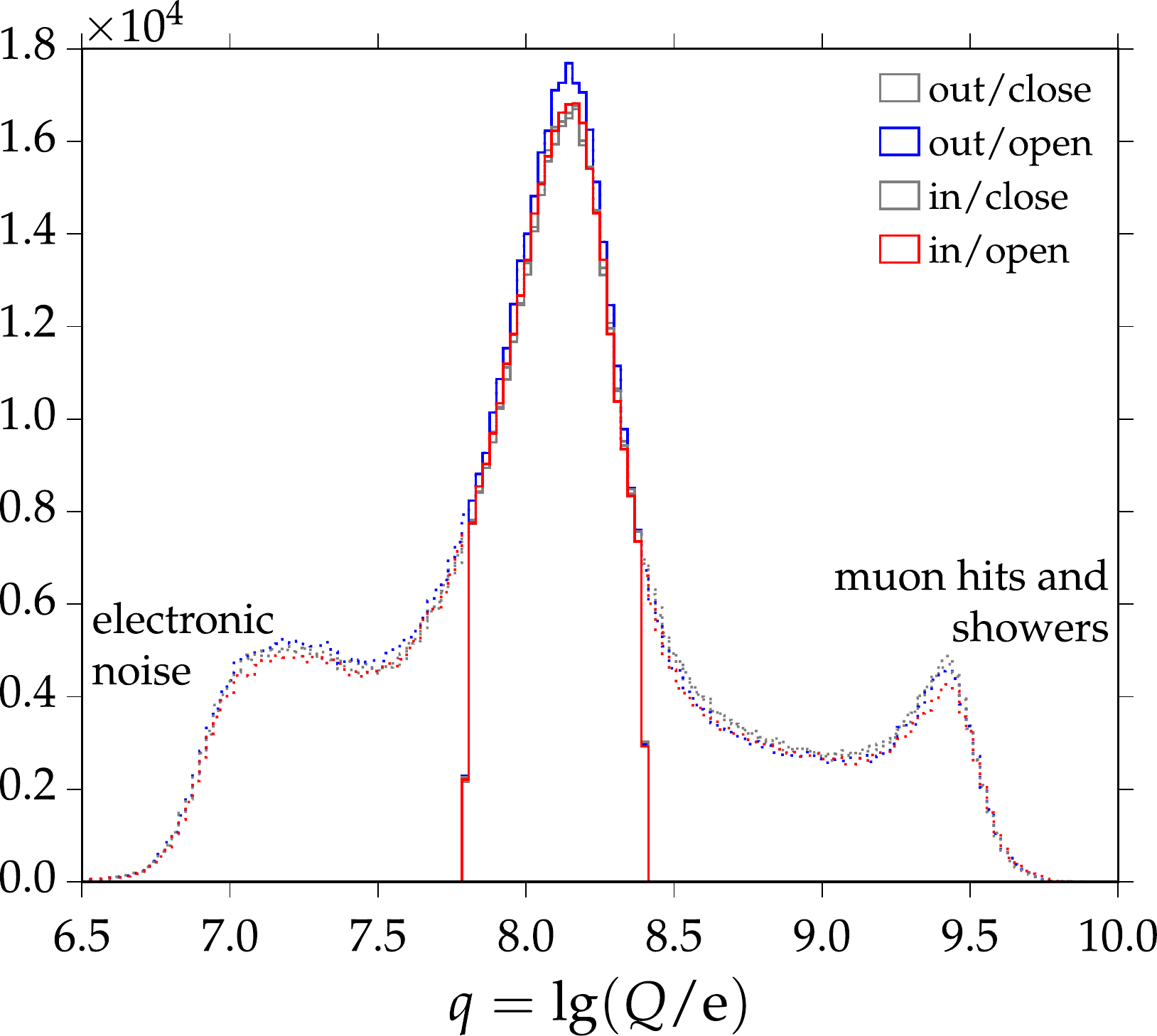}
\caption{\emph{Left:} Charge distribution for a flasher run with very low power
setting and, therefore, composed mostly by single photo-electrons.
\emph{Right:} Charge distribution observed in one of the measurement runs. Note
that in both cases $q=\lg(Q/\text{e})$ where $Q$ is the charge of a pulse.}
\label{f:charge}
\end{figure}

Signals from the PMT were digitized with the PicoScope 6404D digital
oscilloscope.  In Fig.~\ref{f:traces} two examples of triggered traces are
given.  A single-photon (SP) pulse can be observed on the left and a trace
containing several pulses in the 1.6\,$\upmu$s trigger window is shown on the
right. Traces with multiple pulses were discarded since they can be produced
only by cosmic-ray showers. Based on measurements of the SP charge spectrum
with an LED flasher, Fig.~\ref{f:charge}-left, a cut on the allowed range of
observed charges was also applied, as seen in Fig.~\ref{f:charge}-right. The
efficiency of the latter cut on SP traces is estimated to be 75\%.

\section{Preliminary limit on mixing parameter}

\begin{figure}[t]
\def\figh{0.295}
\centering
\includegraphics[height=\figh\textwidth]{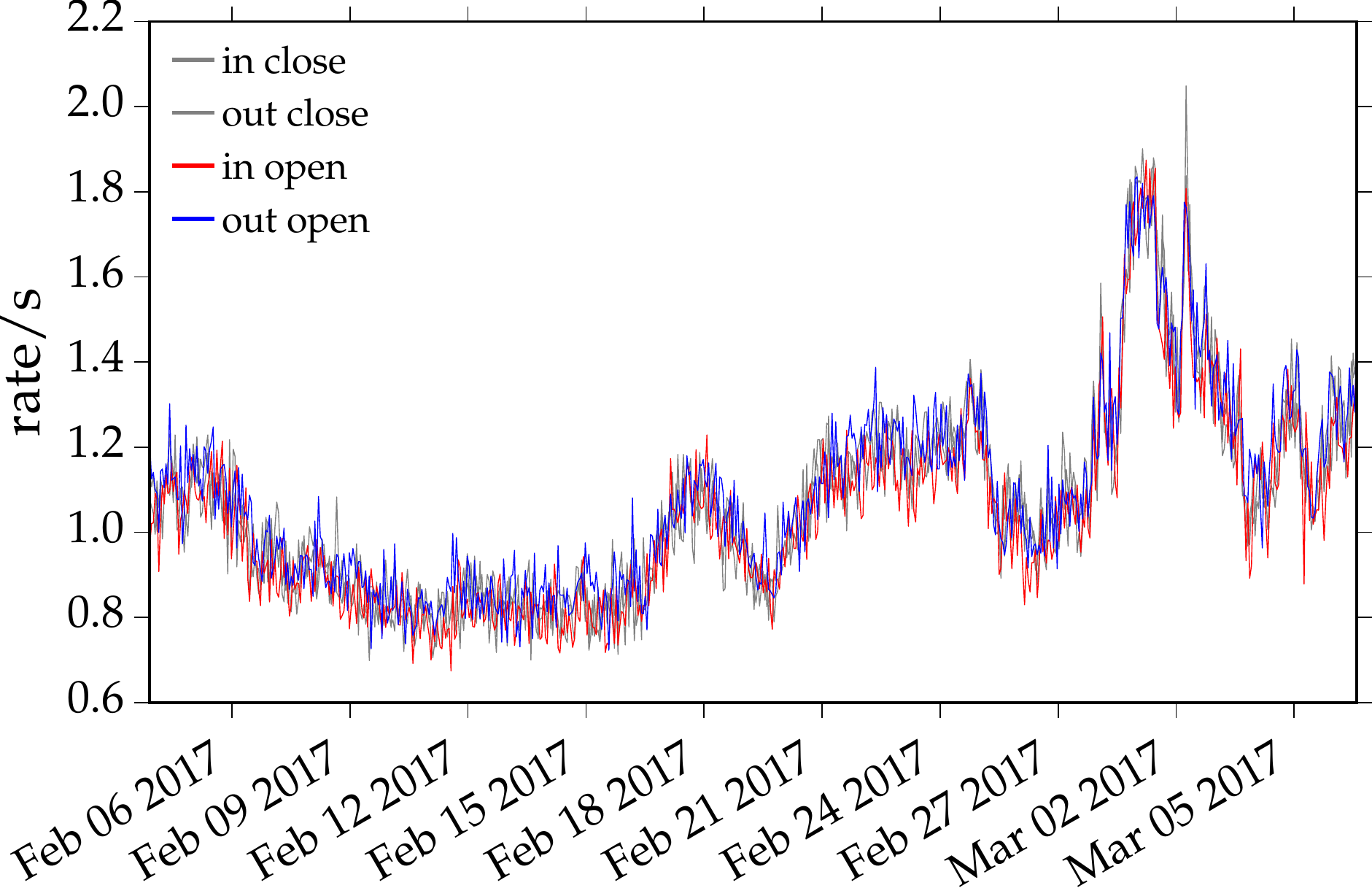}\hfill
\includegraphics[height=\figh\textwidth]{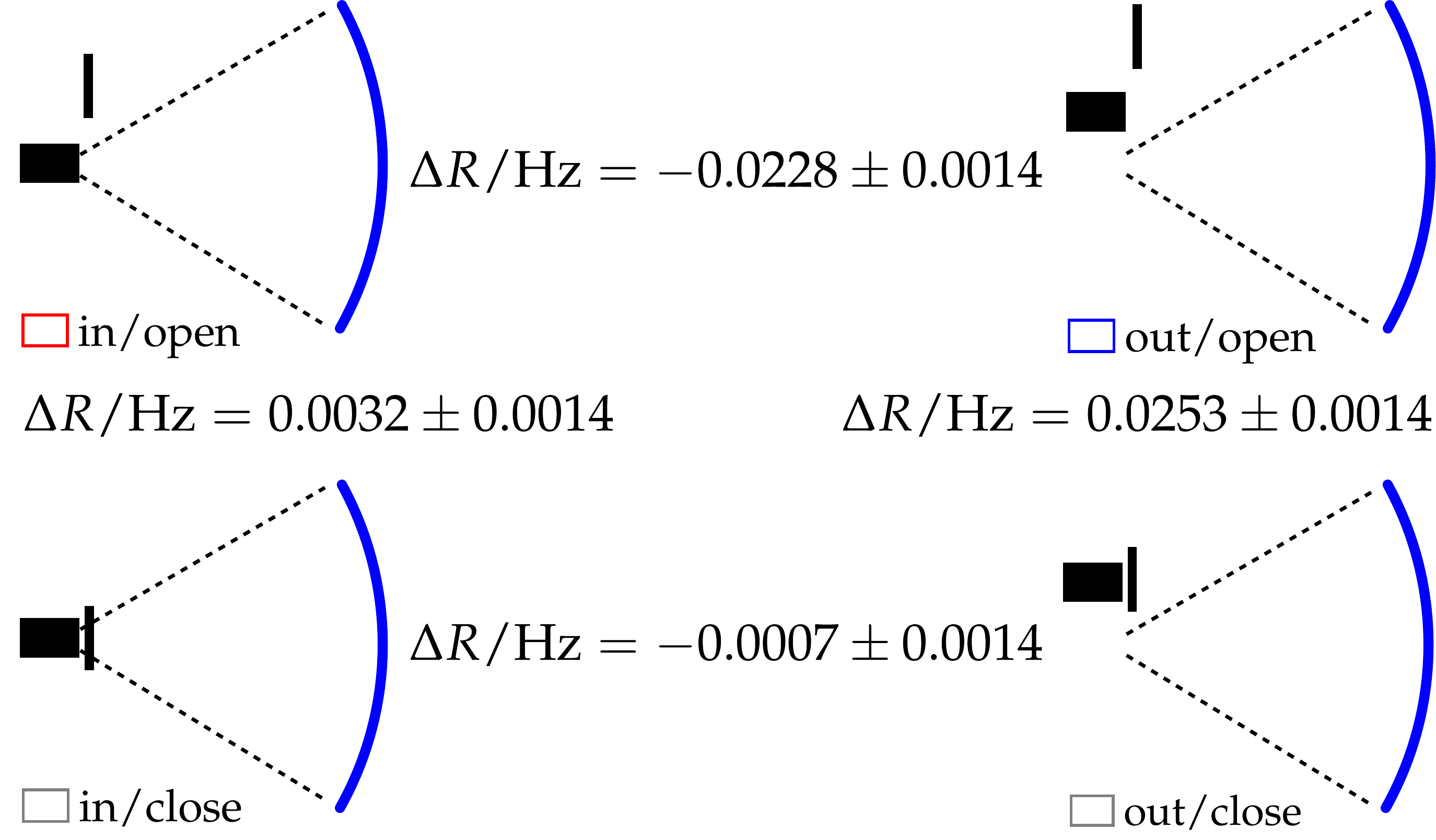}
\caption{\emph{Left:} observed pulse rate in one of the measurement runs.
\emph{Right:} A measurement is composed of many event cycles, where in each
cycle four different 60-second measurements are performed. The schematic show
these four different combinations obtained with open/closed shutter and with
camera in/out.}
\label{f:measurement}
\end{figure}

The selected photon counts for a 30-day run performed in February and March
2017 can be seen in Fig.~\ref{f:measurement}-left.  The data was taken in
cycles of four 60\,s measurements performed in all four possible combinations
of the two positions of the PMT camera (\emph{in} and 8\,cm \emph{out} of the
center), and the two positions of the shutter (\emph{open} and \emph{close}),
as schematically shown in Fig.~\ref{f:measurement}-right. The average rate of
the whole run is $R=0.535$\,Hz and the relative differences $\Delta R$ in the
four different configurations are shown in Fig.~\ref{f:measurement}-right.

\begin{figure}[t]
\def\figh{0.3}
\centering
\includegraphics[height=\figh\textwidth]{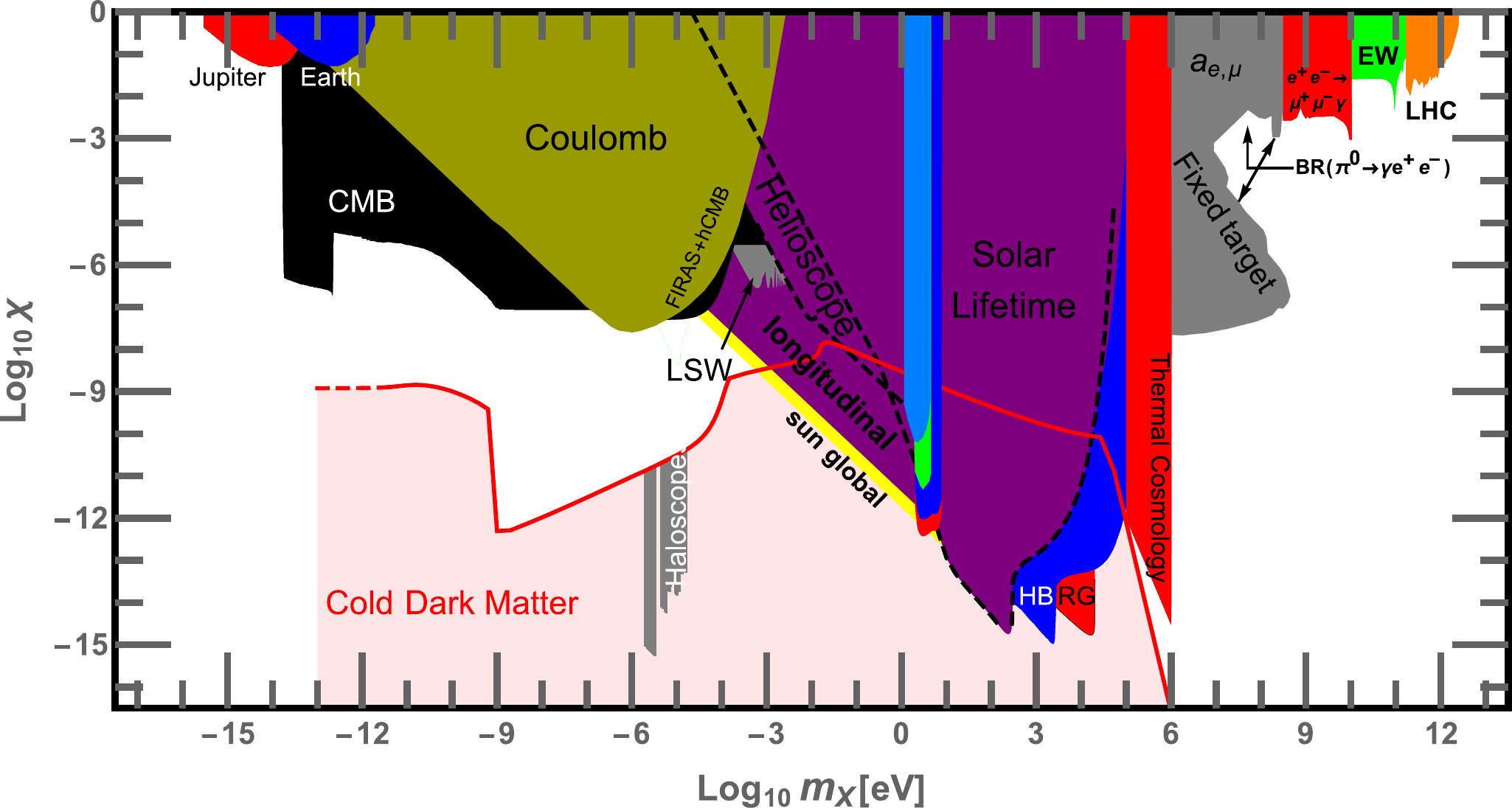}\hfill
\includegraphics[height=\figh\textwidth]{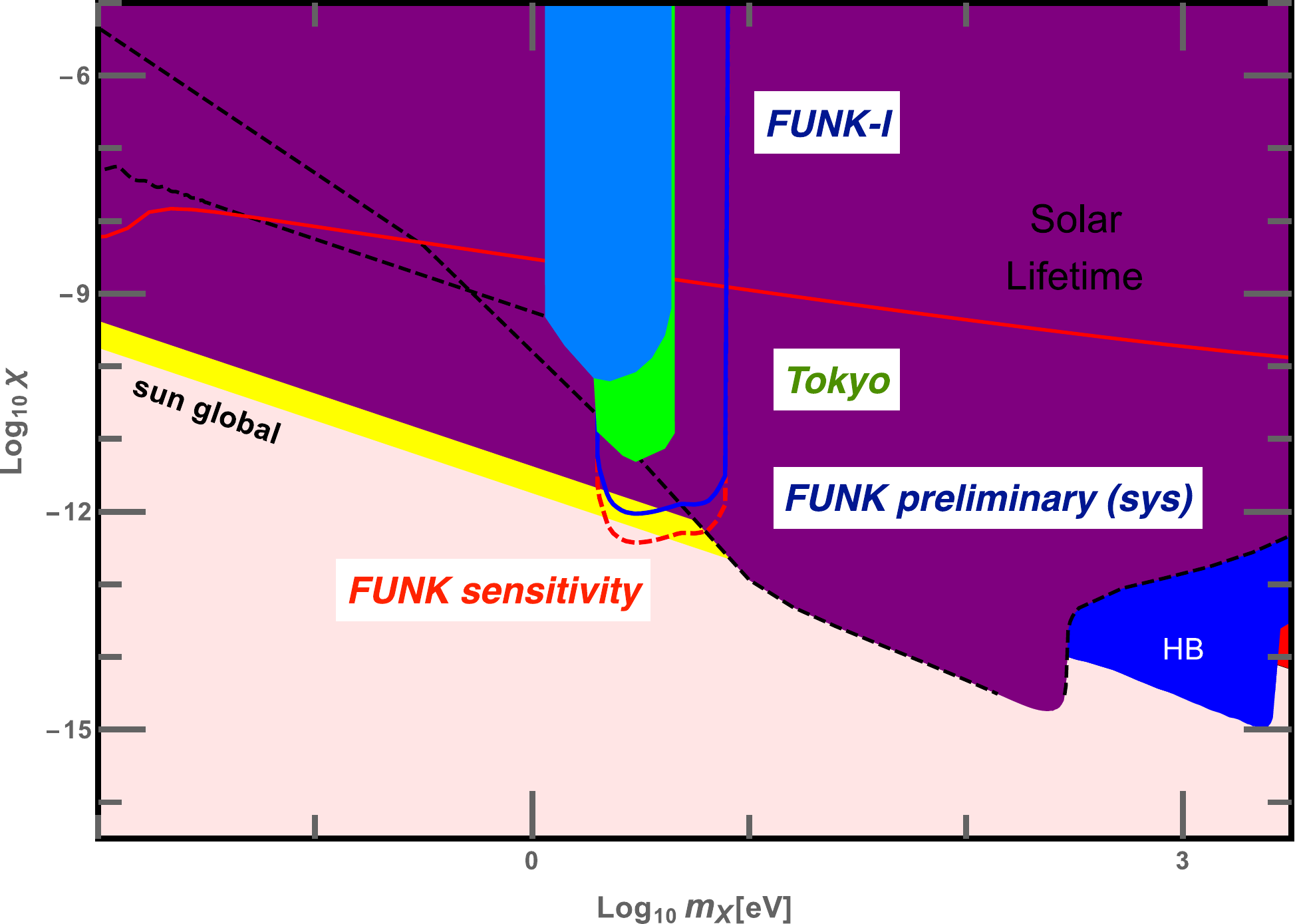}
\caption{Preliminary limits on the mixing parameter $\chi$ derived from
systematic dominated (blue) and statistics dominated (red) assumptions. The
right figure is an enlargement of the right plot around the dark-matter mass
window corresponding to the optical emission of the visible photons. The
previous result obtained with a CCD camera~\cite{Veberic:2015yua} is also shown
(light-blue).}
\label{f:limit}
\end{figure}

The difference of the count rates between \emph{open} and \emph{close} for the
PMT being \emph{in} the radius point is proxy for the dark-matter signal. With
the shutter \emph{open} there are $\Delta R=0.0032\pm0.0014$\,Hz more counts
registered than with \emph{closed}. Ignoring for a moment any possible
systematic effects, we obtain the limit shown in Fig.~\ref{f:limit} denoted
with \emph{FUNK sensitivity}. To determine possible systematic uncertainties
that might be related to temperature changes or the limited accuracy of the
measurement time, we also compare the rates with the \emph{closed} PMT
\emph{in} and \emph{out} of the radius point. The two count rates agree within
the statistical uncertainty ($\Delta R=0.0007\pm0.0014$\,Hz).  Nevertheless,
the comparison of the count rates for the \emph{open} PMT \emph{in} and
\emph{out} of the radius point we found significantly larger count rate for the
PMT \emph{open} and \emph{out} of the radius point, possibly related to the
different imaging properties of the setup in the two positions.  Additional
measurements are in progress to better understand this systematic behavior. For
now we treat this difference ($\Delta R\approx0.025$\,Hz) as an upper limit of
the overall systematic uncertainty of the measurement and use it to derive a
preliminary upper limit~\cite{Dobrich:2014kda} on the magnitude of the mixing
parameter $\chi$ in the sensitivity range of the PMT, see the line denoted
\emph{FUNK preliminary (sys)} in Fig.~\ref{f:limit}.

\textbf{Summary.} No significant signal was found. The detailed analysis of the
data is still ongoing, thus here we are reporting only preliminary results with
a maximally conservative estimate of possible systematic uncertainties.

\textbf{Future plans.} We are planning further searches for possible
hidden-photon dark matter with measurements extended into the MHz, GHz, and THz
range.

\textbf{Acknowledgments.} We gratefully acknowledge partial support from the
Helmholtz Alliance for Astroparticle physics (HAP), funded by the Initiative
and Networking Fund of the Helmholtz Association.

\begin{footnotesize}

\end{footnotesize}

\end{document}